\documentstyle[aps,preprint,tighten]{revtex}

\begin{document}

\draft

\title{Two-dimensional electron gas in a uniform magnetic field 
in the presence of a $\delta$-impurity}
\author{R. M. Cavalcanti\footnote{Present address: Institute for
Theoretical Physics, University of California, Santa Barbara, 
CA 93106-4030, USA. E-mail: rmc@itp.ucsb.edu}}
\address{Departamento de F\'\i sica, Pontif\'\i cia Universidade 
Cat\'olica do Rio de Janeiro \\
CP 38071, CEP 22452-970, Rio de Janeiro, RJ, Brasil}
\author{C. A. A. de Carvalho\footnote{E-mail: aragao@if.ufrj.br}}
\address{Instituto de F\'\i sica, Universidade Federal do 
Rio de Janeiro \\
CP 68528, CEP 21945-970, Rio de Janeiro, RJ, Brasil}
\maketitle

\begin{abstract}

The density of states and the Hall conductivity of a two-dimensional
electron gas in a uniform magnetic field and in the presence
of a $\delta$-impurity are exactly calculated using elementary
field theoretic techniques. The impurity creates one localized
state per Landau level, but the Hall conductivity is unaffected.
Our treatment is explicitly gauge-invariant, and can be easily
adapted to other problems involving zero-range potentials. 

\end{abstract}


\section{Introduction}

One of the most puzzling features of the quantum Hall effect 
is the apparent insensitivity of the quantization
of the Hall conductivity $\sigma_H$ (in multiples of $e^2/h$) with 
respect
to type of host material, geometry of sample, presence of impurities,
etc. Prange\cite{Prange,QHE} was probably the first to address the
question of the influence of impurities on the quantization
of $\sigma_H$. For a two-dimensional electron gas
in crossed electric and magnetic fields in the presence of
a $\delta$-impurity, he showed that a localized state
exists, which carries no current, while the remaining nonlocalized
states carry an extra Hall current which exactly compensates
for the part not carried by the localized state.

Notwithstanding Prange's claim that his calculation was exact, he
had in fact to resort to some approximations. In part, this
occurred because he worked with a finite (although small)
electric field, and in part because he used a $\delta$-function 
potential,
which is too `strong' in two (or more) dimensions\cite{Tarrach}, even
in the presence of a magnetic field.
As shown explicitly by Perez and Coutinho\cite{Perez}, if one
solves the Schr\"odinger equation for a square well of
radius $a$ and depth $V_0(a)\sim a^{-2}$, one finds a
bound state whose energy $E_b\to -\infty$ when $a\to 0$.
In order for $E_b$ to remain finite in the limit $a\to 0$,
the depth of the well must diverge more slowly than $a^{-2}$;
explicitly, $V_0(a)\sim 1/a^2\ln (a/R)$ ($R$ is some constant
with dimension of length). Working with this regularized version
of the $\delta$-function, they managed to find a spectrum similar
to the one found by Prange.

The purpose of this paper is to revisit this problem using
elementary field theoretic techniques. There are a couple of
reasons for doing things this way: (i) it is very easy to find the 
Feynman propagator in the presence 
of a $\delta$-impurity\cite{Economou,Albeverio}, 
and (ii) the density of states and the 
conductivity tensor can be computed exactly, in an
explicitly gauge-invariant way.
The very singular nature of the $\delta$-function potential in two 
dimensions
shows up in our treatment as an infinity in the propagator,
but to deal with it is a very simple exercise in 
renormalization\cite{Albeverio}. 

Our calculations essentially confirm Prange's results.


\section{The Feynman propagator}

Let us consider an electron gas in two dimensions in 
a uniform magnetic field, in the presence of a
$\delta$-function potential at the origin. Its lagrangian density is
given by (we use units such that $m=\hbar=c=1$)
\begin{equation}
{\cal L}={\psi}^{\dag}\,(i\partial_t-H+\mu)\,\psi.
\end{equation}
The `1-particle hamiltonian' $H$ can be split in two pieces:
$H=H_0+V({\bf x})$, where
\begin{equation}
H_0=-\frac{1}{2}\,({\bf \nabla}-ie{\bf A})^2,\qquad
V({\bf x})=\lambda\,\delta^2({\bf x}).
\end{equation}
The vector potential {\bf A} generates a uniform magnetic field
$B$ ($=\partial_1A_2-\partial_2A_1$), and $\mu$ denotes
the chemical potential.
 
As we shall see in the next section, the particle density
and the conductivity of the system can be computed once one
knows the Feynman propagator, which satisfies the following
equation ($x\equiv(t,{\bf x})$):
\begin{equation}
\label{Green}
(i\partial_t-H+\mu)_x\,G(x,x')=\delta^3(x-x').
\end{equation}
Since $H$ is time-independent, we can look for a solution of
(\ref{Green}) in the form
\begin{equation}
\label{Gt}
G(x,x')=\int_{-\infty}^{\infty}\frac{d\omega}{2\pi}\,
e^{-i\omega(t-t')}\,G(\omega;{\bf x},{\bf x}').
\end{equation}
This in fact is a solution of (\ref{Green}), provided
$G(\omega;{\bf x},{\bf x}')$ is a solution of
\begin{equation}
\label{Green2}
(\omega-H+\mu)_x\,G(\omega;{\bf x},{\bf x}')=\delta^2({\bf x}-{\bf 
x}').
\end{equation}
This can be solved in the usual way as
\begin{equation}
\label{Green3}
G(\omega;{\bf x},{\bf x}')=\sum_n\frac{\psi_n({\bf x})\,
\psi_n^*({\bf x}')}{\omega-E_n+\mu},
\end{equation}
where $E_n$ and $\psi_n({\bf x})$ are the eigenvalues and 
eigenfunctions
of $H$, respectively. Since $H$ is a hermitian operator, its 
eigenvalues
are real, so that a prescription must be provided to deal with the 
poles
of $G(\omega;{\bf x},{\bf x}')$ when one performs the integral
over $\omega$ in (\ref{Gt}). For the Feynman propagator, this
amounts\cite{Abrikosov} to deform the integration contour in the
complex $\omega$-plane as indicated in Figure 1.

Defining the `unperturbed' propagator $G_0(\omega;{\bf x},{\bf x}')$
as the solution of (\ref{Green2}) with $V=0$, we can formally solve 
for $G$ as 
\begin{equation}
\label{Dyson}
G=G_0+G_0VG_0+G_0VG_0VG_0+\ldots
\end{equation}
(The product of two operators $F$ and $G$ is defined as 
$(FG)\,({\bf x},{\bf x}')\equiv\int d^2y\,F({\bf x},{\bf y})\,
G({\bf y},{\bf x}')$, and $V({\bf x},{\bf x}')$ stands 
for $V({\bf x})\,\delta^2({\bf x}-{\bf x}')$.)
Because of the very simple form of
$V$, all the integrals in (\ref{Dyson}) can be performed exactly,
and the series can be summed in closed form
(we drop the dependence on $\omega$ for simplicity):
\begin{eqnarray}
G({\bf x},{\bf x}')&=&G_0({\bf x},{\bf x}')+\lambda\,
G_0({\bf x},{\bf 0})\,\sum_{n=0}^{\infty}\left[\lambda\,
G_0({\bf 0},{\bf 0})\right]^n G_0({\bf 0},{\bf x}') \nonumber \\
&=&G_0({\bf x},{\bf x}')+\frac{G_0({\bf x},{\bf 0})\,
G_0({\bf 0},{\bf x}')}{\frac{1}{\lambda}-G_0({\bf 0},{\bf 0})}.
\label{sum}
\end{eqnarray}
One can verify, by direct substitution in Eq.~(\ref{Green2}) (with
$V=0$), that the `unperturbed' Feynman propagator is given by
($eB$ is assumed positive)
\begin{equation}
\label{G0}
G_0(\omega;{\bf x},{\bf x}')=\frac{eB}{2\pi}\,M({\bf x},{\bf x}')\,
e^{-eB({\bf x}-{\bf x}')^2/4}\,\sum_{n=0}^{\infty}
\frac{L_n\left(eB\,({\bf x}-{\bf 
x}')^2/2\right)}{\omega-(n+1/2)\,eB+\mu},
\end{equation}
where $L_n(z)$ is a Laguerre polynomial\cite{GR} and
$M({\bf x},{\bf x}')$ is a gauge dependent factor, which can
be written in a gauge covariant way as
\begin{equation}
M({\bf x},{\bf x}')=\exp\left\{ie\int_{{\bf x}'}^{{\bf x}}
{\bf A}({\bf z})\cdot d{\bf z}\right\},
\end{equation}
the integral being performed along a straight line connecting
${\bf x}'$ to {\bf x}.

Given the explicit form of $G_0(\omega;{\bf x},{\bf x}')$, 
(\ref{sum}) gives the solution of Eq.~(\ref{Green2}), but,
as it stands, it is meaningless: the denominator
of the second term on the r.h.s.\ is logarithmically divergent. 
However, this divergence can be absorbed in a 
redefinition of the `coupling constant' $\lambda$: 
introducing a convergence factor $e^{-\alpha n}$
in the sum over Landau levels, one finds ($z\equiv 
1/2-(\omega+\mu)/eB$)
\begin{eqnarray}
G_0(\omega;{\bf 0},{\bf 0})&=&\frac{eB}{2\pi}\,
\sum_{n=0}^{\infty}\frac{e^{-\alpha n}}{\omega-(n+1/2)\,eB+\mu}
\nonumber \\
&=&-\frac{1}{2\pi}\left[\sum_{n=0}^{\infty}e^{-\alpha 
n}\left(\frac{1}{z+n}
-\frac{1}{n+1}\right)+\sum_{n=0}^{\infty}\frac{e^{-\alpha 
n}}{n+1}\right]
\nonumber \\
&\approx&\frac{1}{2\pi}\,[\gamma+\psi(z)+\ln \alpha]\qquad(\alpha\to 
0^+);
\end{eqnarray}
$\psi(z)$ denotes the digamma function and $\gamma=0.577\ldots$
is Euler's constant\cite{Abramowitz}. Now, we define a renormalized
`coupling constant' $\lambda_R$ as
\begin{equation}
\frac{1}{\lambda_R}=\frac{1}{\lambda}-\frac{1}{2\pi}\,(\gamma+\ln 
\alpha),
\end{equation}
and make $\lambda$ depend on $\alpha$ in such a way that
$\lambda_R$ remains finite\footnote{Note that, for this to be possible, 
$\lambda$ must be negative, i.e., the potential is attractive. 
If one starts with a repulsive 
$\delta$-potential, there is no way to deal with the divergent terms in 
the perturbative
expansion of $G$. This reflects the fact that in
dimensions $D\ge 2$ a repulsive $\delta$-potential expels the $S$-waves
from the Hilbert space\cite{Tarrach}. 
In this paper we consider only the attractive case.}
in the limit $\alpha\to 0^+$.
In terms of the renormalized `coupling constant,' Eq.~(\ref{sum}) 
becomes
\begin{equation}
\label{Green4}
G({\bf x},{\bf x}')=G_0({\bf x},{\bf x}')
+\frac{G_0({\bf x},{\bf 0})\,G_0({\bf 0},{\bf x}')}
{\frac{1}{\lambda_R}-\frac{1}{2\pi}\,\psi\left(\frac{1}{2}-
\frac{\omega+\mu}{eB}\right)}\equiv
G_0({\bf x},{\bf x}')+G_1({\bf x},{\bf x}'),
\end{equation}
which is now well defined. 


\section{Particle density, localized states and Hall conductivity}

\subsection{Particle density}

The particle density is given by\cite{Abrikosov}
\begin{equation}
n(x)=-i\,\lim_{t'\to t+0}\,G(t,{\bf x};t',{\bf x}).
\end{equation}
The `unperturbed' part of $n(x)$ is position independent:
\begin{equation}
\label{n0}
n_0=-i\,\lim_{\alpha\to 0^+}\lim_{\varepsilon\to 0^+}
\int\frac{d\omega}{2\pi}\,e^{i\omega\varepsilon}\,
\frac{eB}{2\pi}\sum_{n=0}^{\infty}\frac{e^{-\alpha n}}
{\omega-(n+1/2)\,eB+\mu}.
\end{equation}
Because of the exponential in front of the sum, one can
close the contour depicted in Fig.~1 with a semicircle
of infinite radius in the upper half-plane,
and use residues to evaluate the
integral. The result, after taking the limits $\varepsilon\to 0^+$
and $\alpha\to 0^+$ (in this order), is
\begin{equation}
n_0=\frac{eB}{2\pi}\sum_{n=0}^{\infty}\theta(\mu-(n+1/2)\,eB),
\end{equation}
where $\theta(x)$ is the Heaviside step function. 

With respect to the `perturbed' part of $n(x)$, it is easier to compute
$N_1\equiv\int d^2x\,n_1(x)$, where $n_1(x)\equiv n(x)-n_0$:
\begin{equation}
N_1=-i\,\lim_{\varepsilon\to 0^+}
\int\frac{d\omega}{2\pi}\,e^{i\omega\varepsilon}
\int d^2x\,\frac{G_0(\omega;{\bf x},{\bf 0})\,
G_0(\omega;{\bf 0},{\bf x})}{\frac{1}{\lambda_R}
-\frac{1}{2\pi}\,\psi\left(\frac{1}{2}-\frac{\omega+\mu}{eB}\right)}.
\end{equation}
Performing the integral over {\bf x} with the help
of the identity $\int_0^{\infty} 
e^{-z}\,L_m(z)\,L_n(z)\,dz=\delta_{m,n}$
one finds:
\begin{equation}
N_1=-\frac{i}{4\pi^2eB}\,\lim_{\varepsilon\to 0^+}
\int d\omega\,e^{i\omega\varepsilon}\,
\frac{\psi'\left(\frac{1}{2}-\frac{\omega+\mu}{eB}\right)}
{\frac{1}{\lambda_R}-\frac{1}{2\pi}\,\psi\left(\frac{1}{2}-
\frac{\omega+\mu}{eB}\right)},
\end{equation}
where $\psi'(z)=d\psi(z)/dz$. The integration over $\omega$
can also be performed using residues, but now there are two
classes of poles to consider. The poles of the first class
have the form $\omega_n^{(1)}=-\mu+(n+1/2)\,eB$ $(n=0,1,2,\ldots)$.
They are second order poles of $\psi'$, but are also simple
poles of $\psi$, and so are simple poles of the integrand.
Their contribution to $N_1$ reads
\begin{equation}
N_1^{(1)}=-\sum_{n=0}^{\infty}\theta(\mu-(n+1/2)\,eB).
\end{equation}
The poles of the second class are given by the roots of the
equation
\begin{equation}
\label{roots}
\frac{1}{\lambda_R}-\frac{1}{2\pi}\,\psi\left(\frac{1}{2}-
\frac{\omega+\mu}{eB}\right)=0.
\end{equation}
Examining the graph of the digamma function\cite{Abramowitz} one 
realizes that the roots of Eq.~(\ref{roots}) have the
form $\omega_n^{(2)}=-\mu+(k_n+1/2)\,eB$, where $k_0<0$ and $n-1<k_n<n$
$(n=1,2,\ldots)$. 
Their contribution to $N_1$ has the opposite sign:
\begin{equation}
N_1^{(2)}=\sum_{n=0}^{\infty}\theta(\mu-(k_n+1/2)\,eB).
\end{equation}
Taking into account an area factor $A$, one finally obtains
the following result for $N\equiv\int d^2x\,n(x)$:
\begin{equation}
\label{N}
N=\left(\frac{eB}{2\pi}\,A-1\right)\sum_{n=0}^{\infty}
\theta(\mu-(n+1/2)\,eB)+\sum_{n=0}^{\infty}\theta(\mu-(k_n+1/2)\,eB).
\end{equation}
This result has a very simple physical interpretation:
since the potential has zero-range, only the $S$-waves are
affected by it. They are expelled from the
Landau levels (which have $eB/2\pi$ states per unit area), 
and mix among themselves to give new states with
energies equal to $(k_n+1/2)\,eB$. The explicit form
of their wave functions are obtained in the next subsection.


\subsection{Localized states}

The wave functions of the localized states can also be obtained from
the Green's function. According to (\ref{Green3}) and (\ref{Green4}),
\begin{eqnarray}
\psi_{\ell}({\bf x})\,\psi_{\ell}^*({\bf x}')&=&
\lim_{\omega\to\omega_{\ell}^{(2)}}\left(\omega-\omega_{\ell}^{(2)}
\right) G(\omega;{\bf x},{\bf x}')
\nonumber \\
&=&\frac{2\pi eB}{\psi'(-k_{\ell})}\,G_0(\omega_{\ell}^{(2)};{\bf 
x},{\bf 0})
\,G_0(\omega_{\ell}^{(2)};{\bf 0},{\bf x}').
\end{eqnarray}
It follows from this and Eq.~(\ref{G0}) that
\begin{equation}
\label{psi_l}
\psi_{\ell}({\bf x})=\sqrt{\frac{eB}{2\pi\,\psi'(-k_{\ell})}}\,
M({\bf x},{\bf 0})\,e^{-eB{\bf x}^2/4}\,\sum_{n=0}^{\infty}
\frac{L_n(eB{\bf x}^2/2)}{k_{\ell}-n}.
\end{equation}
For the lowest energy bound state ($\ell=0$),
the generating function of Laguerre polynomials, Eq.~(\ref{Laguerre}),
allows us to express the sum in (\ref{psi_l}) as an
integral: since $k_0<0$, we may write
\begin{eqnarray}
\sum_{n=0}^{\infty}\frac{L_n(\xi)}{k_0-n}&=&
-\sum_{n=0}^{\infty}L_n(\xi)\,\int_0^{\infty}ds\,e^{(k_0-n)s}
\nonumber \\
&=&-\int_0^{\infty}ds\,e^{k_0 s}\,\sum_{n=0}^{\infty}L_n(\xi)\,
e^{-ns}
\nonumber \\
&=&-\int_0^{\infty}\frac{ds}{1-e^{-s}}\,\exp\left(k_0 s-
\frac{\xi}{e^s -1}\right).
\end{eqnarray}
Changing the variable of integration from $s$ to
$z=\left(e^s -1\right)^{-1}$, we find
\begin{equation}
\label{sumL}
\sum_{n=0}^{\infty}\frac{L_n(\xi)}{k_0-n}=
-\int_0^{\infty} e^{-\xi z}\,z^{-k_0-1}\,(1+z)^{k_0}\,dz
=-\Gamma(-k_0)\,U(-k_0,1,\xi),
\end{equation}
where $U(a,b,z)$ is the Kummer's function which is singular
at the origin\cite{Abramowitz}. 

Combining (\ref{psi_l}) and (\ref{sumL}) we finally find
\begin{equation}
\label{psi_0}
\psi_0({\bf x})=-\sqrt{\frac{eB}{2\pi\,\psi'(-k_0)}}\,\Gamma(-k_0)\,
M({\bf x},{\bf 0})\,e^{-eB{\bf x}^2/4}\,U(-k_0,1,eB{\bf x}^2/2).
\end{equation}
This has precisely the form found by Perez and Coutinho\cite{Perez}
using the method discussed in the Introduction.

Although the condition $k_0<0$ was essential for obtaining
(\ref{sumL}), this expression can be analitically continued
for $k_00$ ($k_0\ne 1,2,\ldots$), thus generalizing (\ref{psi_0})
for the other localized states. Note also that, although they diverge 
at
the origin ($U(a,1,z)\approx -[\,\ln z +\psi(a)]/\Gamma(a)$
when $|z|\to 0$), the wave functions $\psi_{\ell}({\bf x})$ are 
normalizable.


\subsection{Hall conductivity}

Now, let us consider the Hall conductivity. In the linear response
approximation, it is given by\cite{Abrikosov}
\begin{eqnarray}
\label{sigmaij}
\sigma_{21}(x)&=&\frac{e^2}{2}\,\lim_{t'\to t+0}\,
\left(D_{x^{}_2}-D^*_{x'_2}\right)\int d^3y\,G(x,y)\,y_1\,
G(y,x')\,\Big|_{{\bf x}'={\bf x}}
\nonumber \\
&=&\frac{e^2}{2}\,\lim_{\varepsilon\to 0^+}\,
\int\frac{d\omega}{2\pi}\,e^{i\omega\varepsilon}\,
\left(D_{x^{}_2}-D^*_{x'_2}\right)\int d^2y\,G(\omega;{\bf x},{\bf 
y})\,y_1\,
G(\omega;{\bf y},{\bf x}')\,\Big|_{{\bf x}'={\bf x}}\, ,
\end{eqnarray}
where ${\bf D}={\bf \nabla}-ie{\bf A}$ is the gauge covariant 
derivative,
and ${\bf D}^*$ is its complex conjugate. After performing the 
derivatives 
in $x$ and the integration over {\bf y} (for the latter, it
is useful to use the identities\cite{GR} $L_n(z)=L_n^1(z)-L_{n-1}^1(z)$ 
and
$\int_0^{\infty} 
z\,e^{-z}\,L_m^1(z)\,L_n^1(z)\,dz=(n+1)\,\delta_{m,n}$),
the `unperturbed' piece of
the Hall conductivity, obtained by replacing $G$ with $G_0$
in (\ref{sigmaij}), reads 
\begin{equation}
\sigma_{21}^{(00)}=\frac{ie^3B}{8\pi^2}\,\lim_{\alpha\to 0^+}\,
\lim_{\varepsilon\to 0^+}\int d\omega\,e^{i\omega\varepsilon}
\sum_{n=0}^{\infty}[(2n+1)\,f_n^2
-2\,(n+1)\,f_n\,f_{n+1}],
\end{equation}
where
\begin{equation}
f_n\equiv\frac{e^{-\alpha n}}{\omega-(n+1/2)\,eB+\mu}.
\end{equation}
Performing the remaining integral (along the contour of Figure 1),
and taking the limits $\varepsilon,\alpha\to 0^+$, one finally
obtains
\begin{equation}
\label{sigma00}
\sigma_{21}^{(00)}=-\frac{e^2}{2\pi}\,\sum_{n=0}^{\infty}
\theta(\mu-(n+1/2)\,eB).
\end{equation}

The `perturbed' piece of the Hall conductivity, 
obtained by replacing $G$ with $G_1$ in Eq.~(\ref{sigmaij}),
is easily shown to be zero (the integrand is an odd function of $y_1$).
One can also compute exactly the space average of 
the `perturbed-unperturbed' pieces (in which one of the $G$'s 
in Eq.~(\ref{sigmaij}) is replaced by $G_0$ and the other by $G_1$);
the calculation is rather tedious (it is sketched in the Appendix),
but the result is remarkably simple: it is zero. 

Since all that remains is the `unperturbed' piece of $\sigma_{21}$,
we recover the remarkable result of Prange that the Hall
conductivity of a two-dimensional electron gas is not affected by a
$\delta$-impurity, even though such an impurity is
capable of producing localized states.

As a final remark, we notice that there
is a simple reason why the correction to the Hall conductivity
should vanish: as shown by St\v{r}eda\cite{Streda}, when the
chemical potential is in an energy gap the Hall conductivity
is given by following expression:
\begin{equation}
\sigma_{21}=-\frac{ec}{A}\,\frac{\partial N}{\partial B},
\end{equation}
where $N$ is the number of states below the chemical 
potential $\mu$ and $A$ is the area of the system.
With $N$ given by Eq.~(\ref{N}) (and remembering that $c=1$
in our units), it follows from the above expression
and from Eq.~(\ref{sigma00}) that $\sigma_{21}=\sigma_{21}^{(00)}$.

\vspace{1cm}

{\em Note added.\/} The Green's function of a particle in a uniform magnetic
field in the presence of a short-range impurity was previously obtained
by Gesztezy {\em et al} (Gesztezy F, Holden H and \v{S}eba P 1989
On point interactions in magnetic field systems {\em Schr\"odinger
Operators, Standard and Non-standard\/} ed P Exner and P \v{S}eba
(Singapore: World Scientific) pp 147--164). We thank Prof.~Pavel Exner
for calling our attention to that paper.


\acknowledgments

This work had financial support from CNPq, CAPES, FINEP and FUJB/UFRJ.


\appendix

\section{}

In this appendix we sketch the calculation of the spatial average of
$\sigma_{21}^{(01)}(x)$, obtained
by replacing the first $G$ in Eq.~(\ref{sigmaij}) with $G_0$
and the second with $G_1$. An explicit calculation shows that
\begin{eqnarray}
\label{DG}
& &\left(D_{x^{}_2}-D^*_{x'_2}\right)\int d^2y\,
G_0(\omega;{\bf x},{\bf y})\,y_1\,
G_1(\omega;{\bf y},{\bf x}')\,\Big|_{{\bf x}'={\bf x}}
\nonumber \\
& &\qquad\qquad = \frac{eBy_1\,G_0({\bf y},{\bf 
0})}{\frac{1}{\lambda_R}
-\frac{1}{2\pi}\,\psi\left(\frac{1}{2}-\frac{\omega+\mu}{eB}\right)}\,
\bigg\{\left[-ix_1+\frac{iy_1}{2}+\frac{y_2}{2}\right]
G_0({\bf x},{\bf y})\,G_0({\bf 0},{\bf x})
\nonumber \\
& &\qquad\qquad + (x_2-y_2)\,{\cal G}_0({\bf x},{\bf y})\,G_0({\bf 
0},{\bf x})
- x_2\,G_0({\bf x},{\bf y})\,{\cal G}_0({\bf 0},{\bf x})\bigg\},
\end{eqnarray}
where ${\cal G}_0$ is obtained from $G_0$ by replacing the Laguerre
polynomials $L_n(z)$ by their derivatives with respect to the
argument, $L_n'(z)$. It follows from the explicit form
of $G_0$ and ${\cal G}_0$ and from the
identity $L_n'(z)=-L^1_{n-1}(z)$ that, when calculating 
$\sigma_{21}^{(01)}\equiv A^{-1}\int d^2x\,\sigma_{21}^{(01)}(x)$, 
one has to deal with integrals which, after the change of variables
$({\bf x},{\bf y})\to\sqrt{eB/2}\,({\bf x},{\bf y})$ is performed,
have the following form:
\begin{equation}
\label{intxy}
I_{\ell mn}^{\alpha\beta\gamma}[P(x_i,y_j)]\equiv
\int d^2x\,d^2y\,e^{-F({\bf x},{\bf y})}\,
P(x_i,y_j)\,L_{\ell}^{\alpha}({\bf x}^2)\,
L_{m}^{\beta}(({\bf x}-{\bf y})^2)\,
L_{n}^{\gamma}({\bf y}^2),
\end{equation}
where
\begin{equation}
F({\bf x},{\bf y}) = i\,\epsilon_{ij}x_iy_j+\frac{1}{2}\,[
{\bf x}^2+({\bf x}-{\bf y})^2+{\bf y}^2]
\end{equation}
and $P(x_i,y_j)$ is a polynomial (of second degree) in $x_i$, $y_j$. 

With the help of the generating function of Laguerre 
polynomials\cite{GR},
\begin{equation}
\label{Laguerre}
(1-t)^{-1-\alpha}\,\exp\left(\frac{tz}{t-1}\right)=
\sum_{n=0}^{\infty}L_n^{\alpha}(z)\,t^n\qquad(|t|<1),
\end{equation}
we define another generating function:
\begin{eqnarray}
\label{cal Z}
{\cal Z}_{\alpha\beta\gamma}(t,u,v;{\bf p},{\bf q})&\equiv&
\sum_{\ell,m,n=0}^{\infty}I_{\ell mn}^{\alpha\beta\gamma}
[e^{{\bf p}\cdot{\bf x}+{\bf q}\cdot{\bf y}}]\,
t^{\ell}\,u^m\,v^n
\nonumber \\
&=&(1-t)^{-1-\alpha}(1-u)^{-1-\beta}(1-v)^{-1-\gamma}\int d^2x\,d^2y\,
e^{{\bf p}\cdot{\bf x}+{\bf q}\cdot{\bf y}-F({\bf x},{\bf y})}\,
\nonumber \\
& &\times
\exp\left\{\frac{t\,{\bf x}^2}{t-1}+\frac{u\,({\bf x}-{\bf y})^2}{u-1}
+\frac{v\,{\bf y}^2}{v-1}\right\} \qquad(|t|,|u|,|v|<1).
\end{eqnarray}
Performing the integrals over {\bf x} and {\bf y}, we obtain
\begin{equation}
{\cal Z}_{\alpha\beta\gamma}(t,u,v;{\bf p},{\bf q})
=\frac{4\pi^2\eta\,e^{\eta\,(a\,{\bf p}^2+b\,{\bf q}^2+c\,{\bf 
p}\cdot{\bf q}-
i\,\epsilon_{ij}p_iq_j)}}{(1-t)^{1+\alpha}(1-u)^{1+\beta}(1-v)^{1+\gamma}},
\end{equation}
where
\begin{mathletters}
\begin{eqnarray}
a&=&\frac{1-uv}{(1-u)(1-v)}, \\
b&=&\frac{1-tu}{(1-t)(1-u)}, \\
c&=&\frac{1+u}{1-u}, \\
\eta&=&\frac{(1-t)(1-u)(1-v)}{4(1-tuv)}.
\end{eqnarray}
The integrals in (\ref{intxy}) can then be obtained as the coefficients
of the expansion in a power series in $t$, $u$ and $v$ of
\end{mathletters}
\begin{equation}
P(\partial_{p_i},\partial_{q_j})\,
{\cal Z}_{\alpha\beta\gamma}(t,u,v;{\bf p},{\bf q})
\big|_{{\bf p}={\bf q}={\bf 0}}.
\end{equation}

The first type of integrals we need to evaluate is 
$I_{\ell mn}^{000}[-ix_1y_1+(iy_1^2+y_1y_2)/2]$. Following the
recipe given above, we obtain
\begin{eqnarray}
& &\sum_{\ell,m,n=0}^{\infty}I_{\ell mn}^{000}
[-ix_1y_1+(iy_1^2+y_1y_2)/2]\,t^{\ell}\,u^m\,v^n 
\nonumber \\
& &\qquad\qquad=\left(-i\,\frac{\partial^2}{\partial p_1\partial q_1}
+\frac{i}{2}\,\frac{\partial^2}{\partial q_1^2}
+\frac{1}{2}\,\frac{\partial^2}{\partial q_1\partial q_2}
\right){\cal Z}_{000}(t,u,v;{\bf p},{\bf q})\big|_{{\bf p}={\bf q}={\bf 
0}}
\nonumber \\
& &\qquad\qquad=\frac{4\pi^2\eta^2\,(-ic+ib)}{(1-t)(1-u)(1-v)}
=\frac{i\pi^2\,(t-u)(1-v)}{4\,(1-tuv)^2}
\nonumber \\
& &\qquad\qquad=\frac{i\pi^2}{4}\,(t-u)(1-v)\,\sum_{k=1}^{\infty}
(k+1)\,t^k\,u^k\,v^k.
\end{eqnarray}
It follows that
\begin{equation}
I_{\ell mn}^{000}[-ix_1y_1+(iy_1^2+y_1y_2)/2]=
\frac{i\pi^2}{4}\,[(m+1)\,\delta_{\ell,m+1}\,(\delta_{n,m}-
\delta_{n,m+1})-(m\leftrightarrow\ell)],
\end{equation}
so that ($f_k\equiv[\,\omega-(k+1/2)\,eB+\mu]^{-1}$)
\begin{equation}
\label{000}
\sum_{\ell,m,n=0}^{\infty}I_{\ell mn}^{000}
[-ix_1y_1+(iy_1^2+y_1y_2)/2]\,f_{\ell}\,f_m\,f_n = 0.
\end{equation}
The other two types of integrals we need can be found in an
analogous way: 
\begin{eqnarray}
I_{\ell mn}^{010}[x_2y_1-y_1y_2]&=&\frac{i\pi^2}{4}\,
(m+1)(\delta_{\ell,m}-\delta_{\ell,m+1})(\delta_{n,m}-\delta_{n,m+1}), 
\\
I_{\ell mn}^{100}[-x_2y_1]&=&-\frac{i\pi^2}{4}\,
(\ell+1)(\delta_{m,\ell}-\delta_{m,\ell+1})(\delta_{n,\ell}-\delta_{n,\ell+1}).
\end{eqnarray}
Therefore,
\begin{equation}
\label{010+100}
\sum_{\ell,m,n=0}^{\infty}\left(I_{\ell,m-1,n}^{010}[x_2y_1-y_1y_2]
+I_{\ell-1,m,n}^{100}[-x_2y_1]\right)f_{\ell}\,f_m\,f_n = 0.
\end{equation}
It follows from (\ref{DG}), (\ref{000}), (\ref{010+100}) and
the explicit form of $G_0$ and ${\cal G}_0$ that
$\sigma_{21}^{(01)}=0$. An analogous calculation shows that
$\sigma_{21}^{(10)}=0$, too.



\newpage

\noindent
\underline{\bf Figure Captions}:

\vspace{5mm}
\noindent
{\bf Figure 1}: Integration contour in the complex $\omega$-plane used
in the definition of the Feynman propagator.

\end{document}